\documentclass{sf2a-conf2014}
\usepackage{graphicx}
\usepackage{hyperref}
\usepackage[]{natbib}  
\usepackage{epstopdf}

\def\BibTeX{{\rm B\kern-.05em{\sc i\kern-.025em b}\kern-.08em
    T\kern-.1667em\lower.7ex\hbox{E}\kern-.125emX}}
\bibpunct{(}{)}{;}{a}{}{,}  

\begin{document}

\TitreGlobal{SF2A 2014}


\title{A search for Vega-like fields in OB stars}

\runningtitle{Vega-like fields in OB stars}

\author{C. Neiner}\address{LESIA, Observatoire de Paris, CNRS UMR 8109, UPMC,
Universit\'e Paris Diderot, 5 place Jules Janssen, 92190 Meudon, France}

\author{C.~P. Folsom$^{2,}$}\address{IRAP, CNRS/Universit\'e de Toulouse, 14 avenue E. Belin, 31400 Toulouse, France}\address{IPAG, UJF-Grenoble 1/CNRS-INSU, UMR 5274, 38041 Grenoble, France}

\author{A. Blazere$^{1,2}$}

\setcounter{page}{237}


\maketitle


\begin{abstract}
Very weak magnetic fields (with a longitudinal component below 1 Gauss) have
recently been discovered in the A star Vega as well as in a few Am stars.
According to fossil field scenarios, such weak fields should also exist in more
massive stars. In the framework of the ANR project Imagine, we have started to
investigate the existence of this new class of very weakly magnetic stars among
O and B stars thanks to ultra-deep spectropolarimetric observations. The first
results and future plans are presented.
\end{abstract}



\section{Introduction}

The magnetic fields present in OBA stars are of fossil origin, i.e. they result
from the seed field present in the molecular cloud from which the star was
formed, rather than being produced by a currently active dynamo like in the Sun.
This original field may have also been enhanced during the early phases of the
life of the star, when it was fully convective, however such a dynamo early in
the star's life is no longer active.

This fossil magnetic field relaxes onto a stable oblique mainly dipolar field
detectable at the stellar surface \citep{duez2010}. According to observations,
this happens in $\sim$10\% of the OBA stars \citep[e.g.][]{wade2014_mimes}.
Fossil field scenarios predict that the remaining $\sim$90\% of OBA stars should
host very weak fields, either because of a bifurcation between stable and
unstable large-scale magnetic configurations in differentially rotating stars
\citep{lignieres2014} or because those 90\% of stars did not reach a stable
configuration yet \citep{braithwaite2013}. Such very weak fields were recently
discovered in some A stars: Vega \citep{lignieres2009}, Sirius
\citep{petit2011}, and a few Am stars (see Blazere et al., these proceedings).
They are called ``Vega-like'' fields. 

In the frame of the ANR Imagine project, we aim to test the existence of such
very weak fields in more massive (OB) stars.

\section{First results}

\begin{figure}[ht!]
 \centering
 \includegraphics[width=0.6\textwidth,clip]{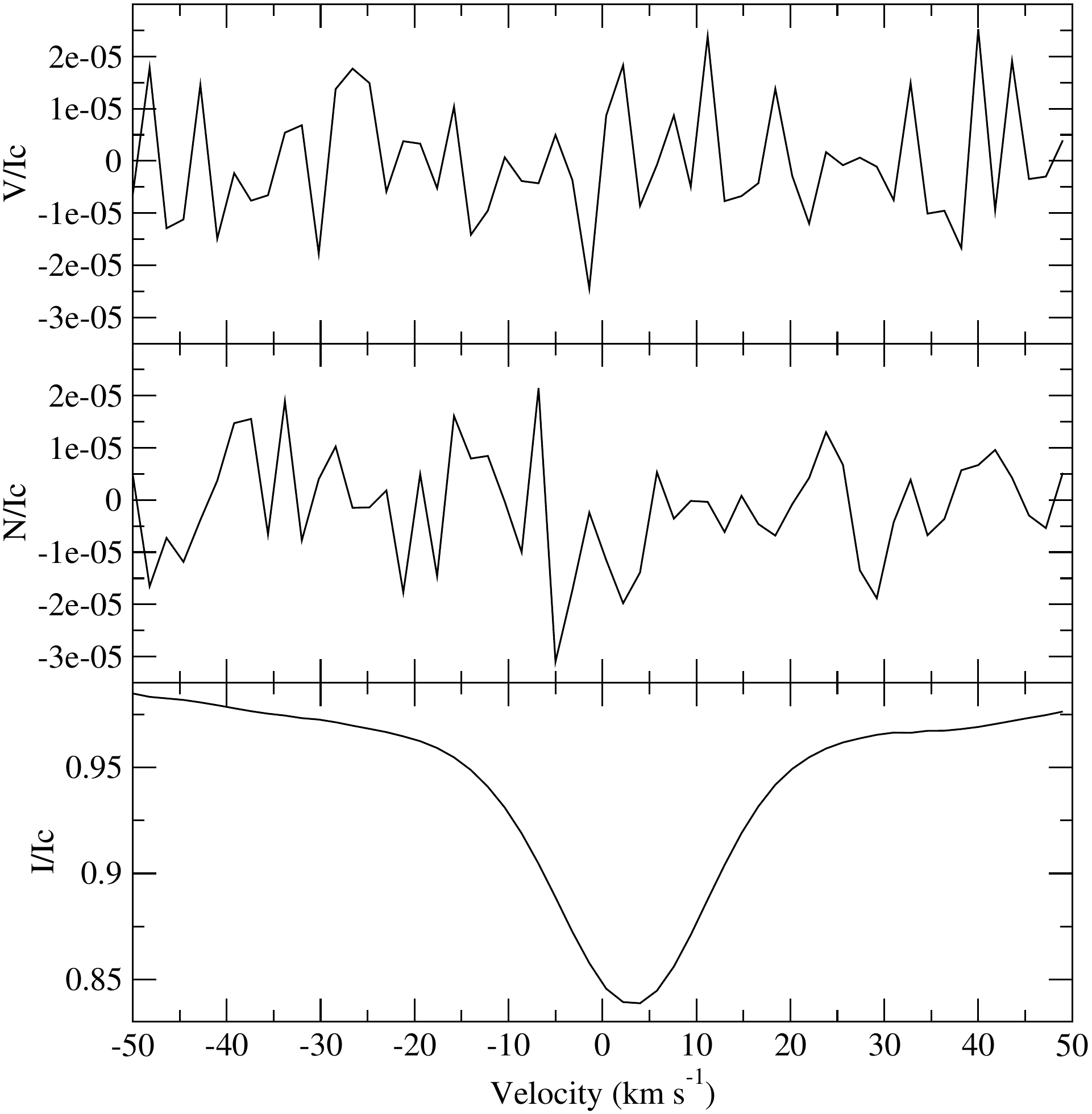}      
  \caption{LSD Stokes V (top), Null polarisation (middle) and I profiles
(bottom) of $\gamma$\,Peg. Stokes V shows no magnetic signature. Taken from
\cite{neiner2014}.}
\label{gampegLSD}
\end{figure}

We have accumulated high resolution, high signal-to-noise spectropolarimetric
Narval observations of the bright slowly rotating B2 star $\gamma$\, Peg. We
analysed these observations using the Least-Squares Deconvolution (LSD)
technique \citep{donati1997} to derive magnetic Zeeman signatures in spectral
lines and the longitudinal magnetic field. With a Monte Carlo simulation we
derived the maximum strength of the field possibly hosted by $\gamma$\,Peg. We
found that no magnetic signatures are visible in the very high quality
spectropolarimetric data. The average longitudinal field measured in the Narval
data is B$_l = -0.1 \pm 0.4$ G \citep[see Fig.~\ref{gampegLSD};][]{neiner2014}.
The precision we reached is thus similar to the one used for the field detection
in A  stars. We derive a very strict upper limit on the strength of any dipolar
field possibly hidden in the noise of our data of B$_{\rm pol} \sim$ 40 G. 

A similar study on the B3 star $\iota$\,Her was performed with high resolution,
high signal-to-noise spectropolarimetric ESPaDOnS data  \citep{wade2014}. The
same analysis technique was used. No Zeeman signatures were detected in the
Stokes V profiles. The longitudinal magnetic field in the average profile was
measured to be B$_l = -0.2 \pm 0.3$ G. Again, the precision is similar to
the one used for the field detection in A stars. An upper limit of the
dipolar field strength of B$_{\rm pol} \sim$ 8 G is derived by \cite{wade2014}.
However, note that their method for extracting dipolar field upper limits is
less conservative than the one we used for $\gamma$\,Peg. 

From these two studies we conclude that no magnetic field is detected in either 
of the two B stars, despite having reached a precision of the longitudinal
magnetic field measurement similar to the one used to detect very weak fields in
A stars. 

\section{Detectability of Vega-like fields in OB stars}

To check the detectability of Vega-like fields in OB stars, we can compare the
observations of $\gamma$\,Peg and $\iota$\,Her with synthetic Stokes V profiles
corresponding to the surface magnetic field strength and geometry of Vega, but
computed for the spectral characteristics of $\gamma$ Peg and $\iota$ Her.  For
this we used the magnetic maps of Vega from \citep{petit2010}, with model line
parameters and $v \sin i$ corresponding to our observations of $\gamma$\,Peg and
$\iota$\,Her. Since the inclination of the rotation axis is not known for either
star, two test values were used.

\begin{figure}[!ht]
 \centering
 \includegraphics[width=0.331\textwidth,clip]{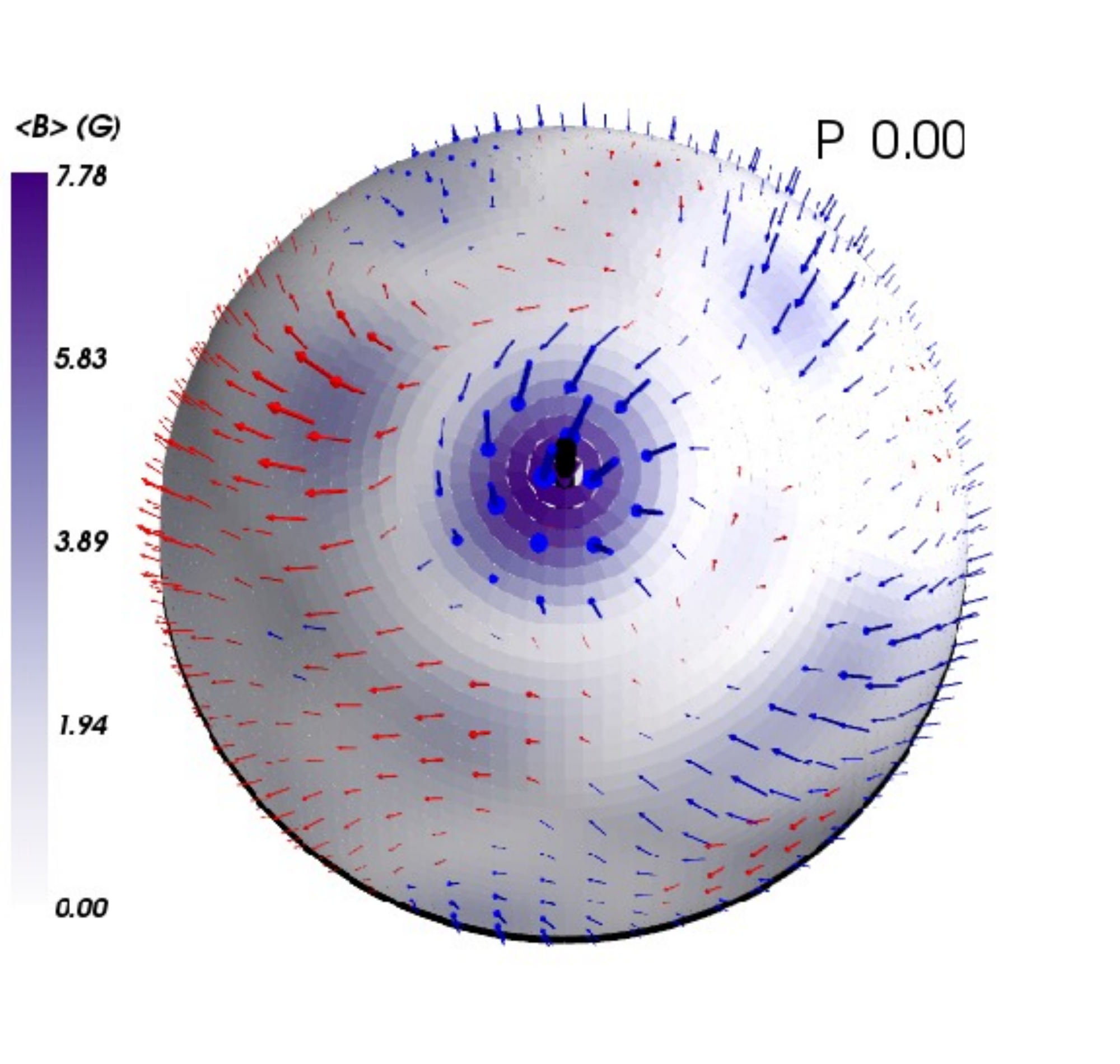}%
 \includegraphics[width=0.331\textwidth,clip]{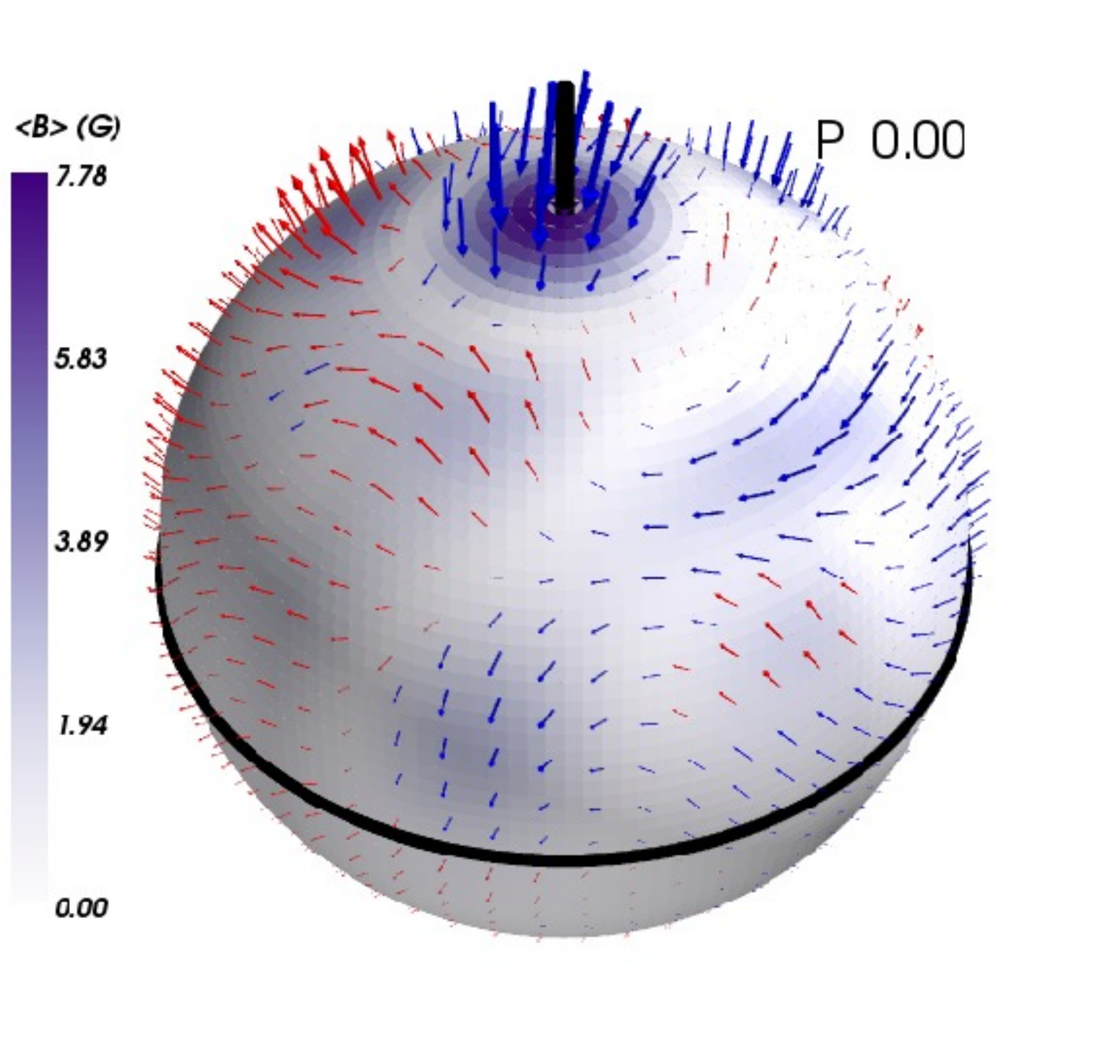}\\      
 \includegraphics[width=0.331\textwidth,clip]{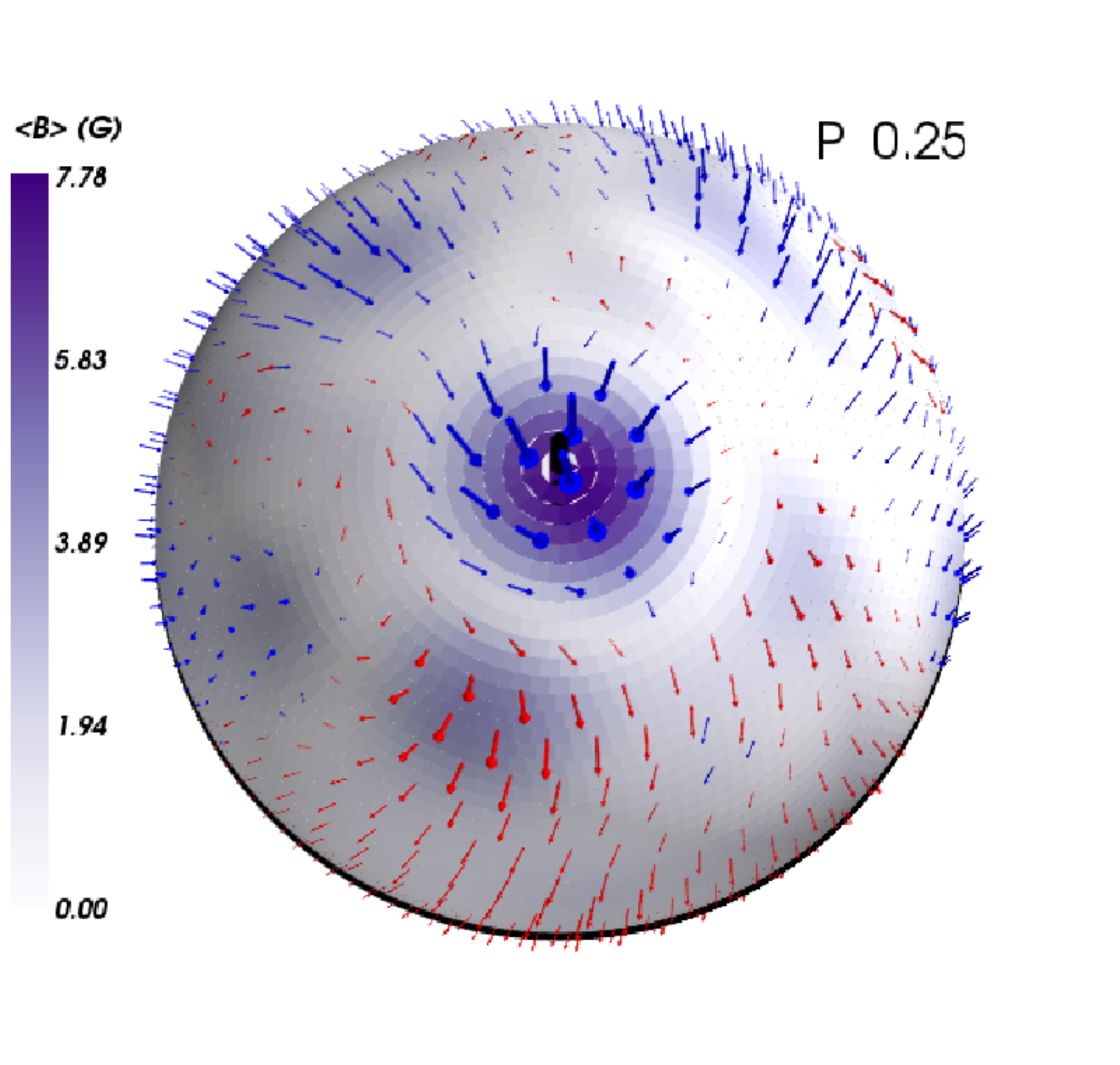}%
 \includegraphics[width=0.331\textwidth,clip]{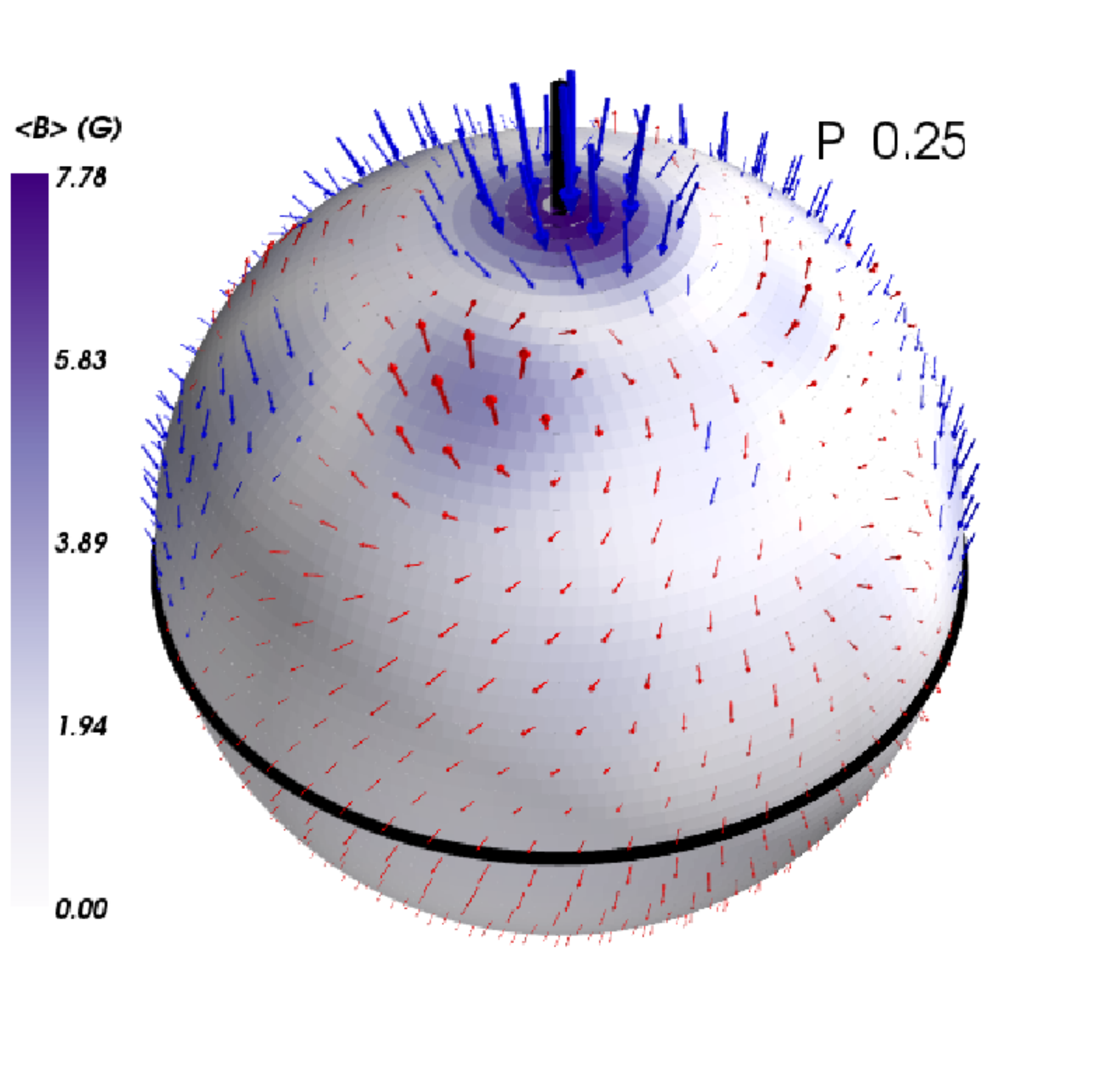}\\      
 \includegraphics[width=0.331\textwidth,clip]{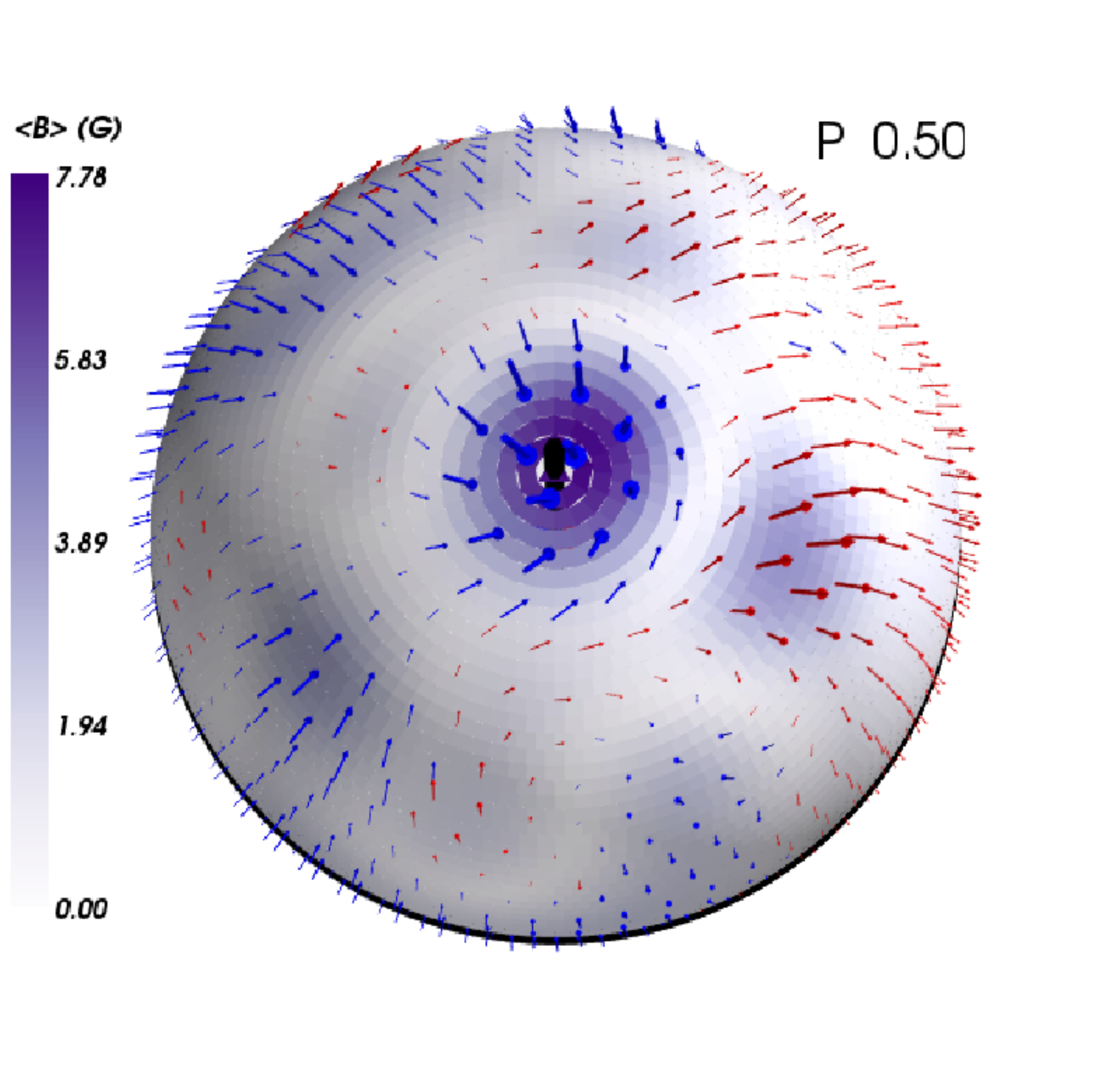}%
 \includegraphics[width=0.331\textwidth,clip]{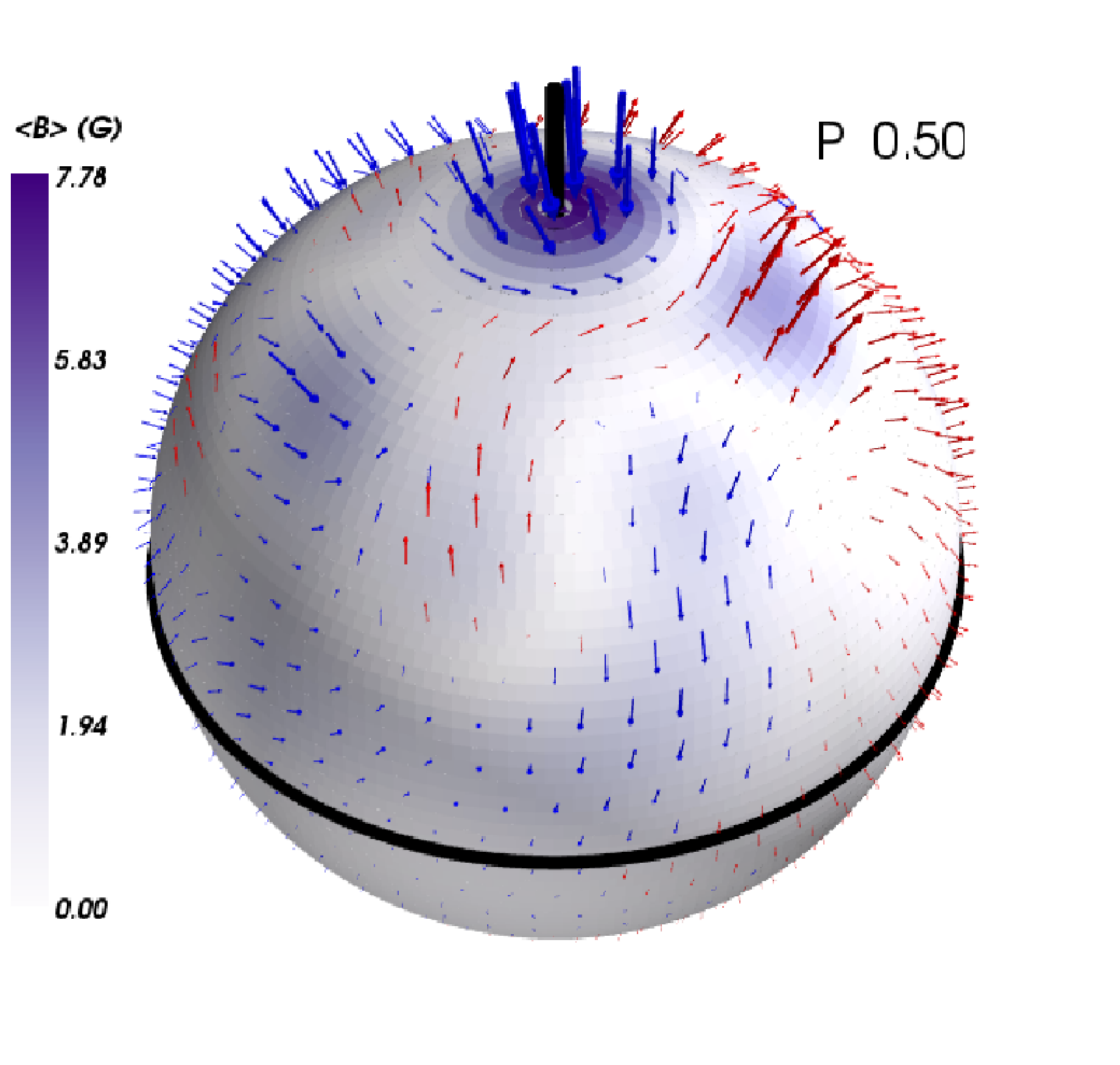}\\      
 \includegraphics[width=0.331\textwidth,clip]{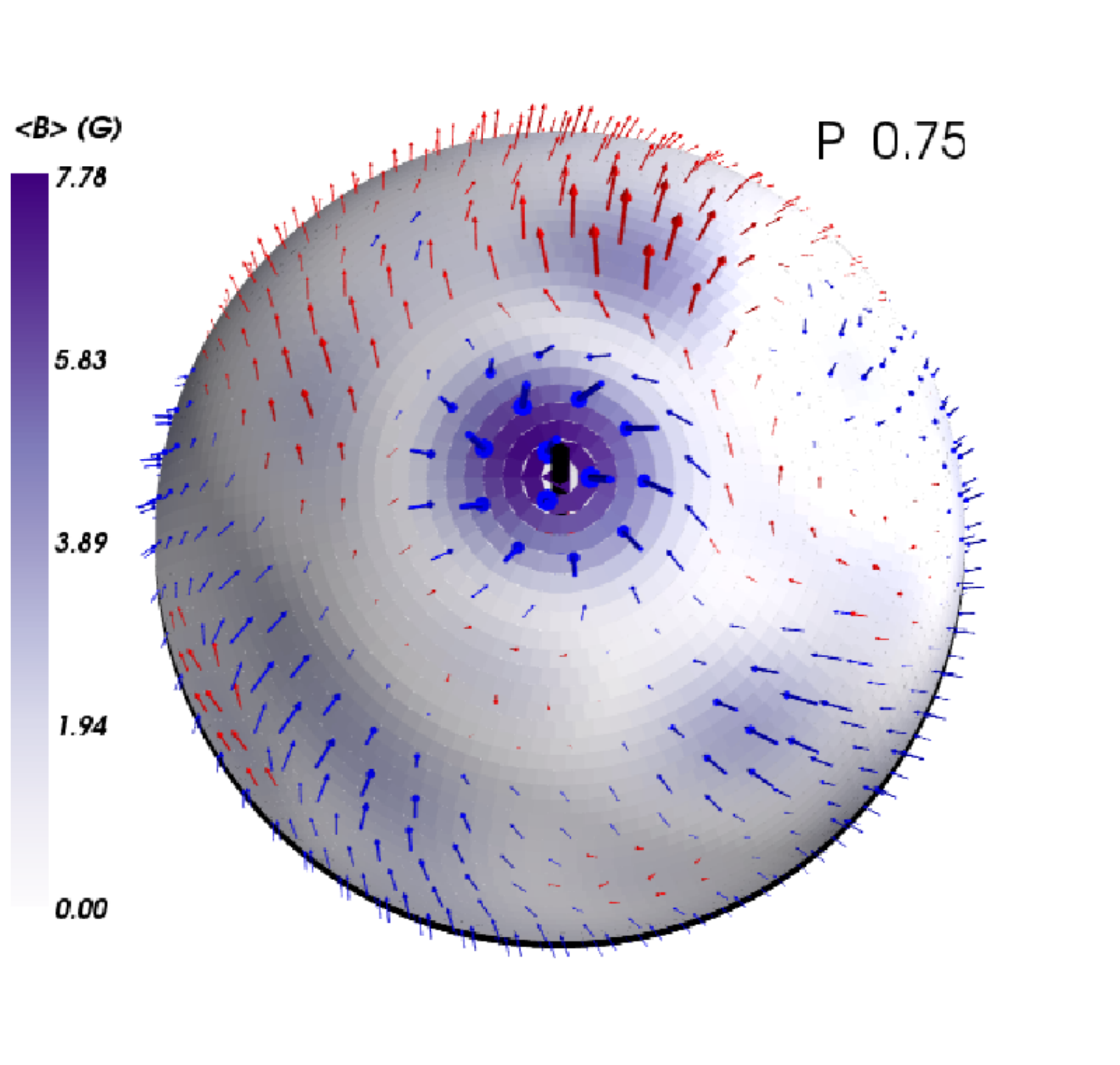}%
 \includegraphics[width=0.331\textwidth,clip]{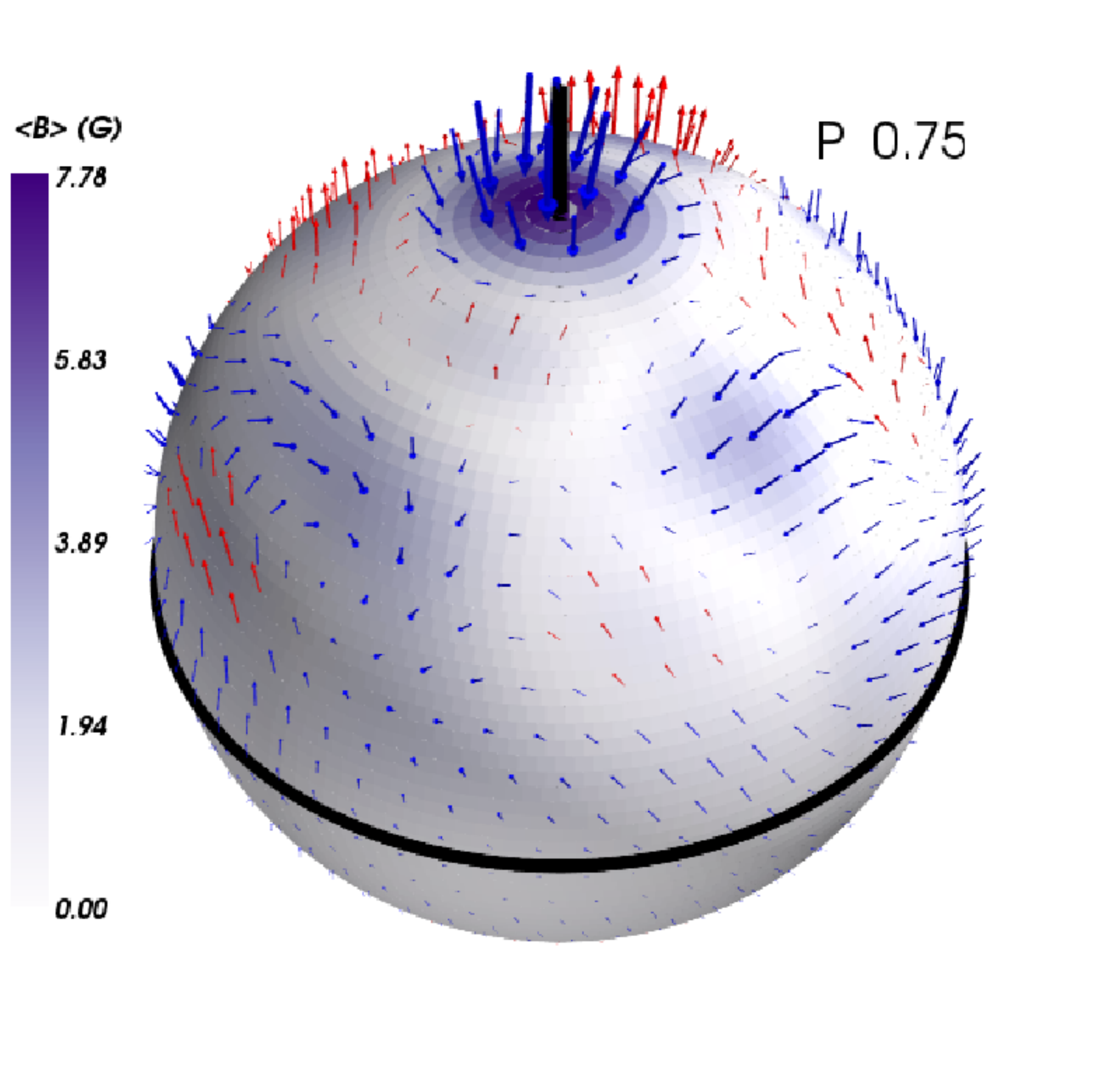}      
  \caption{Maps of the field of Vega applied to $\gamma$\,Peg as seen at various
phases (top to bottom) and under two different inclination angles. {\bf Left:}
i=7$^\circ$ as for Vega. {\bf Right:} i=45$^\circ$.}
  \label{gampegMaps}
\end{figure}

\begin{figure}[ht!]
 \centering
 \includegraphics[width=0.6\textwidth,clip]{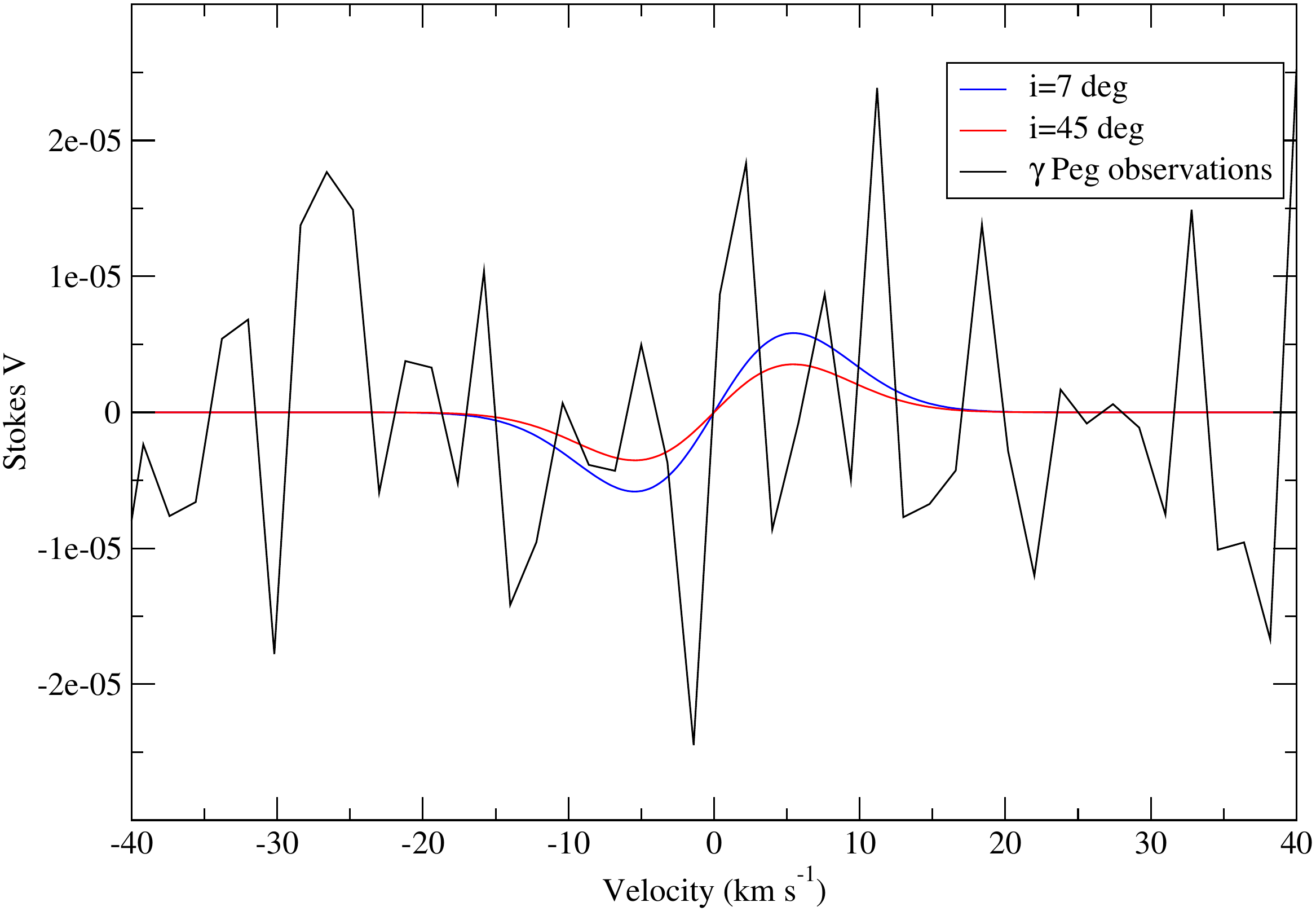}      
  \caption{Expected (rotationally averaged) Stokes V profile if $\gamma$ Peg hosted a magnetic field identical to that of Vega, for two different inclination angles, compared to the mean observed profile.}
\label{gampegModel}
\end{figure}

The results for $\iota$ Her were presented in \cite{wade2014}. They conclude
that is is unlikely a magnetic field identical to Vega's field would have been
detected, unless the observations were made from a particularly favorable
angle.  However, they also conclude that a magnetic field with the same geometry
but $\sim$4 times stronger would almost certainly have been detected.

Here we present the maps of the magnetic field (Fig.~\ref{gampegMaps}) and
Stokes V line profiles (Fig.~\ref{gampegModel}) we would have observed if
$\gamma$\,Peg hosted the same field as Vega, either with the same inclination
angle as Vega ($i=7^\circ$), or with $i=45^\circ$.  Since the rotation period of
$\gamma$\,Peg is unknown, and our observations (distributed over one month)
likely cover several rotational cycles, we consider a rotationally averaged
model line profile.  For $\gamma$\,Peg we find similar results to $\iota$\,Her:
a magnetic field identical to Vega's field is unlikely to have been detected,
but one $\sim$4 times stronger would have likely been detected.

We conclude that, to detect a field like the one of Vega but on an early B star,
we would need to measure longitudinal fields with a precision of 0.1 G (rather
than 0.3-0.4 G reached for $\iota$\,Her and $\gamma$\,Peg and sufficient for A
stars). While it is unlikely we would have detected a magnetic field identical
to Vega's field on $\gamma$\,Peg or $\iota$\,Her, we would have likely detected
a field with a peak strength approximately four times as strong as that of Vega.
The precision required for O stars is probably even higher.

\section{Future work and conclusions}

To investigate the presence of Vega-like fields in OB stars, it will be
necessary to reach a precision on the longitudinal field measurements of at
least 0.1 G. This will require the co-addition of many Narval, ESpaDOnS or
HarpsPol observations to reach a huge  signal-to-noise, i.e. it will require a
large amount of telescope time for each target. We will thus very carefully
select the best O and B targets and propose observations of only a few optimal
stars.

In this way we will test the existence of a new class  of very weakly magnetic
stars, currently only observed among A stars, and characterise it. The existence
of this new class of objects among all OBA stars would lead to a revolution
similar to the one we underwent 15 years ago with the discovery of intermediate
fields in OB stars.

\begin{acknowledgements}
This work is supported by the ANR Imagine project. This research has made use of
the SIMBAD database operated at CDS, Strasbourg (France), and of NASA's
Astrophysics Data System (ADS).
\end{acknowledgements}

\bibliographystyle{aa}  
\bibliography{Neiner1} 

\end{document}